# Dirac fermions at high-index surfaces of bismuth chalcogenide topological insulator nanostructures

Naunidh Virk[1] and Oleg V. Yazyev[1,*]

[1]*Institute of Theoretical Physics, Ecole Polytechnique Fédérale de Lausanne (EPFL), CH-1015 Lausanne, Switzerland*

Binary bismuth chalcogenides $Bi_2Se_3$, $Bi_2Te_3$, and related materials are currently being extensively investigated as the reference topological insulators (TIs) due to their simple surface-state band dispersion (single Dirac cone) and relatively large bulk band gaps. Nanostructures of TIs are of particular interest as an increased surface-to-volume ratio enhances the contribution of surfaces states, meaning they are promising candidates for potential device applications. So far, the vast majority of research efforts have focused on the low-energy (0001) surfaces, which correspond to natural cleavage planes in these layered materials. However, the surfaces of low-dimensional nanostructures (nanoplatelets, nanowires, nanoribbons) inevitably involve higher-index facets. We perform a systematic *ab initio* investigation of the surfaces of bismuth chalcogenide TI nanostructures characterized by different crystallographic orientations, atomic structures and stoichiometric compositions. We find several stable terminations of high-index surfaces, which can be realized at different values of the chemical potential of one of constituent elements. For the uniquely defined stoichiometric termination, the topological Dirac fermion states are shown to be strongly anisotropic with a clear dependence of Fermi velocities and spin polarization on the surface orientation. Self-doping effects and the presence of topologically trivial mid-gap states are found to characterize the non-stoichiometric surfaces. The results of our study pave the way towards experimental control of topologically protected surface states in bismuth chalcogenide nanostructures.

---

[*] Correspondence and requests for materials should be addressed to O.V.Y. (email: oleg.yazyev@epfl.ch).

Topological insulators (TIs) are a recently discovered class of materials that realizes a non-trivial band structure topology induced by strong spin-orbit interactions.[1,2,3] These novel materials possess topologically protected metallic surface states characterized by robust spin-momentum locking. Initial theoretical predictions[4,5,6,7] and subsequent experimental realizations of bulk TIs[8] have generated an enormous degree of interest in exploring this exotic electronic phase. Significant attention has focused on the so-called "second generation" of bulk TIs – layered binary bismuth chalcogenides $Bi_2X_3$ (X = Se or Te).[9,10] The simple dispersion of their surface-state bands, represented by a single Dirac cone, and relatively large band gaps (~0.3 eV for $Bi_2Se_3$), indicating room temperature stability of the topological phase, have made these materials the "reference" TIs. Despite significant progress in probing surface states via surface sensitive techniques such as angle-resolved photoemission spectroscopy[9,11,12] (ARPES) and scanning tunnelling microscopy/spectroscopy[13,14,15] (STM/STS), electron transport measurements have proven to be more challenging. This is primarily due to surface-state charge carriers being masked by bulk states as a result of intrinsic impurities, such as vacancy and antisite defects,[16,17] or the ambient environment.[18,19]

An efficient strategy towards mitigating bulk contributions is to form low-dimensional nanostructures with increased surface-to-volume ratio.[20,21,22,23,24] Synthesized bismuth chalcogenide nanostructures display a diverse array of morphologies such as hexagonal and triangular platelets, nanowires and nanoribbons, and both linear and irregularly shaped nanoribbons (for example see Ref. 20). In addition to giving access to transport measurements of topologically protected surface states,[25,26,27,28] low-dimensional TI nanostructures also provide the necessary configuration for observing a variety of physical phenomena, for example the Aharonov-Bohm interference.[29,30,31] Finally, one can anticipate novel physical phenomena involving topologically protected charge carriers subjected to reduced dimensions, such as the realization of Majorana fermion quasiparticles.[32]

Bismuth chalcogenides are layered materials composed of covalently bonded sheets termed quintuple layers (QL) (Fig. 1a-1c). These QL building blocks are held together by weak van der Waals (vdW) interactions, thus providing natural cleavage planes and giving rise to two equivalent orientations of low-energy surfaces, (0001) and (000$\bar{1}$). The vast majority of investigations of the surface states of bismuth chalcogenide TIs thus far have focused on the low-energy (0001) surfaces. However, it has to be appreciated that any bismuth chalcogenide nanostructure of dimensionality lower than 2 inevitably exhibits surfaces with orientations other than the two abovementioned. The properties of topologically protected charge carriers at such surfaces would depend on their crystallographic orientation, atomic structure and chemical composition. Furthermore, the properties of a TI nanostructure as a whole are defined by the relative presence of different facets on the nanostructure's surface, which is in turn determined by their respective surface energies.

Fully realizing the potential of bismuth chalcogenide nanostructures for exploring the fundamental physics and device applications of TIs depends on a clear understanding of how structure and morphology of the surfaces impact upon the electronic properties of those nanostructures. So far, only one example of high-index surface of bismuth chalcogenides has been addressed theoretically[33] and experimentally.[34,35] In this paper, we report a systematic first-principles investigation of high-index surfaces of $Bi_2Se_3$ and $Bi_2Te_3$. We first determine the structure and chemical composition of stable surfaces. Subsequently, the electronic properties of corresponding topological surface-state charge carriers and their dependence on surface orientation and local chemical composition are investigated.

## Results

**Stable quintuple-layer terminations and surface energies.** Any investigation of the electronic structure of a material's surface requires detailed knowledge of the atomic structure of stable configurations of that surface. In layered materials, such as $Bi_2Se_3$ and $Bi_2Te_3$, the surface structure is defined by the termination of individual layers, and by their orientation with respect to the surface plane, which we here define in terms of the angle $\theta$, as shown in Figure 1c. In this definition, $\theta = 0°$ corresponds to the degenerate case of the low-energy (0001) surface, for which no QL termination takes place. Furthermore, given that $Bi_2Se_3$ and $Bi_2Te_3$ are layered vdW materials, the surface energy $E$ (expressed in eV/Å$^2$) can subsequently, to a good degree of accuracy, be related to the termination energy of individual quintuple layers $G$ (expressed in eV/Å) as

$$E = \frac{3}{c\sin\theta}G, \qquad (1)$$

where $c$ is the lattice constant of a hexagonal unit cell of the bismuth chalcogenide material, and hence $c/3$ is the QL thickness (9.546 Å for $Bi_2Se_3$ and 10.162 Å for $Bi_2Te_3$[36]). The approximation in Equation 1 stems from the omission of a term related to the loss of vdW energy upon stacking QLs to form the 2D surface. This contribution is assumed to be negligible as the relative magnitude of vdW interactions is small and the leading term to the surface energy is the QL termination energy. Below, as we will be referring to the hexagonal unit cell of bismuth chalcogenides, correspondingly, crystallographic planes and lattice directions will be given in the four-index Miller-Bravais notation. One has to appreciate the fact that QL terminations can have different local stoichiometries, either Bi-rich or Se(Te)-rich, and that their relative stabilities depend on the chemical potential of one of the constituent elements, $\mu_{Bi}$ or equivalently $\mu_{Se}$ ($\mu_{Te}$). The chemical potential reflects experimental conditions under which the nanostructure is grown, and, importantly, its varying values may result in different stable surface structures. Hence, it is possible to tailor the structure of high-index surfaces by changing the experimental conditions, which consequently translates into control over the electronic properties of topologically protected states hosted by these surfaces.

The energies of QL terminations have been systematically investigated by means of density functional theory (DFT) calculations carried out on a large number of single-QL models in a nanoribbon configuration (see Methods). Relative stabilities of different QL terminations were determined by comparing the QL termination free energies $G(\mu_{Bi})$, which are calculated per unit length as a function of the chemical potential of Bi, $\mu_{Bi}$, via the expression

$$G(\mu_{Bi}) = G(\mu_{Bi} = 0) - \frac{N_{Bi}}{2L}\mu_{Bi} \tag{2}$$

with $G(\mu_{Bi} = 0)$ being the QL termination free energy at a reference chemical potential

$$G(\mu_{Bi} = 0) = \frac{1}{2L}(E_{model} - N_{Bi_2X_3}E_{Bi_2X_3} - N_{Bi}E_{Bi}), \tag{3}$$

where X corresponds to Se (Te). Here, $G(\mu_{Bi} = 0)$ is calculated as the formation energy of a QL termination defined as the difference in total energies, per unit length, of a nanoribbon model ($E_{model}$), isolated two-dimensional QL per Bi$_2$Se$_3$ (Bi$_2$Te$_3$) unit ($E_{Bi_2X_3}$), and Bi atom in its bulk elemental crystal ($E_{Bi}$). The reference chemical potential $\mu_{Bi}$ = 0 eV below corresponds to bulk elemental bismuth. $N_{Bi}$ refers to the number of excess Bi atoms in the model relative to a stoichiometric system (i.e. with Bi : X = 2 : 3), while $N_{Bi_2X_3}$ is the number of stoichiometric $Bi_2X_3$ units. *L* is the periodicity of the nanoribbon model, equal to the lattice constant *a* of hexagonal unit cell (4.178 Å for $Bi_2Se_3$ and 4.403 Å for $Bi_2Te_3$, which correspond to values determined here from first-principles) in the case of the QL edge oriented along the $[2\bar{1}\bar{1}0]$ direction, and to $L = \sqrt{3}a$ in the case of the QL edge oriented along the $[1\bar{1}00]$ direction. The $[2\bar{1}\bar{1}0]$ and $[1\bar{1}00]$ lattice directions are labelled in reference to the hexagonal unit cell in Fig. 1b.

The calculated QL termination free energies $G(\mu_{Bi})$ in $Bi_2Se_3$ and $Bi_2Te_3$ are presented in Figures 2a and 2b, respectively. The dependence of $G(\mu_{Bi})$ on the chemical potential $\mu_{Bi}$ is linear, with the slope reflecting the local deviation from the stoichiometric ratio Bi : X = 2 : 3 at the QL edge. Thicker coloured lines indicate stable terminations, i.e. those having the lowest value of $G(\mu_{Bi})$ within a certain range of $\mu_{Bi}$. For $Bi_2Se_3$, we found 7 such configurations: 1 stoichiometric (labelled I in Fig. 2a), 3 Se-rich ($II_{Se}$, $III_{Se}$ and $IV_{Se}$) and 3 Bi-rich ($II_{Bi}$, $III_{Bi}$ and $IV_{Bi}$). Their atomic structure configurations are schematically shown in Figure 2c. The uniquely defined stoichiometric termination, characterized by the constant $G$ = 0.169 eV/Å, is stable in a particularly wide range of $\mu_{Bi}$ values, and it displays only very minor structural relaxation (see Fig. S1 of the Supplementary Information). Several distinct non-stoichiometric terminations are possible because of the varying deviations from ideal stoichiometry indicated in Figure 2c. We note that the stable terminations tend to be oriented along the $[2\bar{1}\bar{1}0]$ direction, with only the $II_{Se}$ configuration being oriented along the $[1\bar{1}00]$ direction. This implies that high-index surfaces formed by the $[2\bar{1}\bar{1}0]$ QL terminations are preferred from the thermodynamic point of view. On the other hand, an observation of a surface formed by the $[1\bar{1}00]$ QL termination unambiguously points to the structure and stoichiometry of the $II_{Se}$ configuration. Analysis of the relaxed atomic structures (Fig. S1 of the Supplementary Information) suggests that the non-stoichiometric configurations with the largest deviations from the stoichiometric ratio, $IV_{Bi}$ and $IV_{Se}$, can be interpreted as the stoichiometric edge configurations I capped by elemental Bi and Se, respectively. This reflects the expected tendency towards the onset of phase segregation into $Bi_2Se_3$ and elemental Bi and Se, respectively, at the extreme values of chemical potential $\mu_{Bi}$. For $Bi_2Te_3$, we predict very similar structures of stable configurations and their associated energies (Fig. 2b). For instance, the stable stoichiometric configuration I with $G$ = 0.161 eV/Å also exists in a broad range of $\mu_{Bi}$ values. However, we find that the QL termination analogous to $III_{Se}$ is not present on its respective phase diagram.

The knowledge of surface energies $E$ for different crystallographic orientations of the surface allows one to deduce the equilibrium crystal shape by means of the Wulff construction.[37] In other words, it enables the contributions of different facets to the overall surface of the nanostructure to be determined. The surface energies of high-index facets can be deduced from the QL termination energies discussed above, except for the (0001) orientation, which is of the vdW origin. We estimate the (0001) surface energy as $E_{vdW} = 0.01$ eV/Å$^2$, using the recently suggested universal interlayer binding energy of layered materials,[38] and a first-principles study of the role of vdW interactions in the binding energy of $Bi_2Se_3$ and $Bi_2Te_3$.[39] This value can be compared with energies of high-index surfaces for $\theta = 90°$ by plotting $G_{equiv} = E_{vdW} c/3$ (dashed line in Figs. 2a,b). In the range of chemical potentials where the most stable QL termination is the stoichiometric configuration I, the low-energy (0001) surface will be dominant. Assuming thermodynamic equilibrium conditions of nanoparticle growth, this will lead to the nanoplatelet or nanoribbon morphology with top and bottom facets being the (0001) and (000$\bar{1}$) low-energy surfaces, while the side surfaces would have the stoichiometric composition with the QL terminations corresponding to configuration I (Fig. 2d). Such nanoplatelets of bismuth chalcogenides have been extensively documented in experimental publications.[21,22,28] For chemical potential values corresponding to Bi-rich or Se(Te)-rich conditions, the energies of stable non-stoichiometric surfaces appear to be lower than those of the (0001) and configuration I stoichiometric surfaces. This implies the nanowire morphology for nanostructures[20,31] grown under such conditions, with the surface dominated by the non-stoichiometric compositions and oriented along the (0001) direction, for example the hexagonal cross-section nanowires shown schematically on the right and left hand side of Fig. 2d. From a general perspective, this points to a possibility of tailoring the morphology, composition and properties of bismuth chalcogenide nanostructures by controlling the growth conditions.

**Topologically protected states at stoichiometric high-index surfaces.** We now turn our discussion to the electronic structure of high-index surfaces of bismuth chalcogenide TIs. As discussed above, the atomic structure of such surfaces is determined by the crystallographic orientation of the surface and the termination of individual QLs. Initially, we focus on the first degree of freedom assuming the stoichiometric termination I, which is stable in wide ranges of chemical potentials in both materials (Figs. 2a–2c). Calculations have been performed on two-dimensional slab models in which the interior part reproduces bulk crystal structure (see Methods). One example is shown in Fig. 1c, and corresponds to $\theta = 57.7°$ and $\theta = 58.0°$ for $Bi_2Se_3$ and $Bi_2Te_3$, respectively, where $\theta$ defines the QL stacking angle of a high-index surface. In this particular surface configuration, atomic planes along which the QLs are terminated almost coincide with the surface place. Figures 3a and 3b show the electronic band structures of $Bi_2Se_3$ and $Bi_2Te_3$ slabs, of 5.4 nm and 5.7 nm width, respectively, with such surface terminations. The band structures are plotted along the *k*-point path with *Γ−A* and *Γ−Y* oriented along the reciprocal lattice vectors associated with the real-space lattice vectors of the surface. Both band structures display a single Dirac cone at the *Γ* point characteristic of topologically protected surfaces states in bismuth chalcodgenide TIs. Additionally, the plots indicate the inverse participation ratio (IPR) of electronic states $\psi_{i,k}(\mathbf{r})$ calculated as

$$\text{IPR}_{i,k} = \int |\psi_{i,k}(\mathbf{r})|^4 \, d\mathbf{r}, \tag{4}$$

which is computed for each state *i* at momentum *k*. Large magnitude of this quantity reflects the localized character of topological surface states. The band structures also show the presence of a gap opening at the Dirac point (3 meV and 4 meV for the $Bi_2Se_3$ and $Bi_2Te_3$ slab models, respectively), as a result of hybridization between surface states localized at opposite surfaces.[40,41,42,43] However, the important observation is the significant anisotropy, i.e. an orientation-dependent band dispersion of surface-localized Dirac fermion states, which agrees well with the results of previous calculations performed on this particular surface[33] – the only high-index surface of $Bi_2Se_3$ addressed theoretically. More importantly, this is also the only case of a high-index surface that was investigated experimentally using samples of $Bi_2Se_3$ epitaxially grown on the InP(001) substrate.[34,35] The degree of Dirac fermion anisotropy, quantified as the ratio of Fermi velocities for the momenta along *x'* direction (i.e. along the QL planes, see Fig. 1c) and along *y'* direction (perpendicular to *x'*), $v_F^{x'}/v_F^{y'} = 2.07$, agrees well with $v_F^{x'}/v_F^{y'} \sim 2$ observed in the ARPES measurements of Ref. 34. Interestingly, the same surface configuration of $Bi_2Te_3$ shows an even more pronounced anisotropy of the Dirac fermion surface states with practically flat bands for the momenta along *y'* direction (Fig. 3c).

Band structures have been calculated for other crystallographic orientations of the surface within the range $25° < \theta < 155°$. Three representative surface configurations are shown in Fig. 4a. For all studied surface orientations the band structures feature an anisotropic Dirac cone at the *Γ* point, but the degree of anisotropy varies across the investigated range of *θ*. Figure 4b shows Fermi velocities computed above the Dirac point energy and for momenta along *x'* and *y'* as a function of surface orientation *θ*, for both $Bi_2Se_3$ and $Bi_2Te_3$. One can notice the following systematic trends in the above-mentioned quantities. Firstly, Fermi velocities of the topological-state bands of high-index surfaces are generally lower than that for the (0001) surface ($4.8\times10^5$ m/s and $4.6\times10^5$ m/s for $Bi_2Se_3$ and $Bi_2Te_3$, respectively, obtained from our calculations). Secondly, the largest anisotropies are achieved around $\theta = \frac{\pi}{2}$, which corresponds to QL planes oriented perpendicular to the surface. In the case of $Bi_2Se_3$, the value of $v_F^{x'}/v_F^{y'}$ does not exceed 2.5, while for $Bi_2Te_3$ the surface-state band is practically dispersionless along *y'* in a broad range of values of *θ* resulting in much larger anisotropies.

Spin texture is another important property of topologically protected surface states. This property can be analysed by investigating the momentum dependence of the expectation values of spin operators

$$\langle S_\alpha(\mathbf{k})\rangle = \frac{\hbar}{2}\langle \psi(\mathbf{k})|\sigma_\alpha|\psi(\mathbf{k})\rangle \quad (\alpha = x,y,z), \tag{5}$$

where $\psi(\mathbf{k})$ are the two-component spinor wavefunctions, and $\sigma_\alpha$ are the Pauli matrices. For all investigated models of high-index stoichiometric surfaces we observe spin-momentum locking of the Dirac fermion surfaces states with a clockwise direction of rotation of the electron spin above the Dirac point. This picture is consistent with the spin texture of topological surface states at the (0001) surface.[41] However, the anisotropy of Dirac fermions manifests itself in the magnitude of the spin-polarization vector $\mathbf{P}(\mathbf{k}) = (2/\hbar)\mathbf{S}(\mathbf{k})$, which is strongly reduced with respect to the nominal value of 1 due to the strong spin-orbit entanglement in bismuth chalcogenide TIs (~0.6 for the (0001) surfaces of $Bi_2Se_3$ and $Bi_2Te_3$[41]). Figure 4c shows the magnitude of the spin-polarization vector **P** of the surface-state charge carriers above the Dirac point energy as a function of surface orientation *θ* for $Bi_2Se_3$ and $Bi_2Te_3$. The magnitude of spin-polarization **P** is always lower along the *x'* direction, with the largest difference achieved for surface orientations around $\theta = \frac{\pi}{2}$, similar to the case of the Fermi velocities. In contrast, the average magnitude of spin-polarization is about the same as for the (0001) surface. This can be compared to the model theory results of Refs. 44,45, which predict that at $\theta = \frac{\pi}{2}$ the spin texture collapses to a single dimension along the $x'$ direction, leading to a complete suppression of spin polarization for momenta oriented along this direction. In our first-principles calculations, however, we do not observe such a vanishing spin polarization, nonetheless its magnitude for momenta along $x'$ direction achieves its minimum down to ~0.4 for both chalcogenide materials.

**Electronic structure of non-stoichiometric high-index surfaces.** The models of non-stoichiometric surfaces are found to reveal generally more complex band structures. In particular, in addition to the topologically protected crossing within the band gap, we observe the presence of topologically trivial mid-gap states and self-doping, i.e. charge transfer between the surface and bulk-like states that leads to the shift of the Fermi level into the valence and conduction bands. Both effects can be considered as detrimental to the observation of topological surfaces states in TI nanostructures. Figure 5 shows two representative examples of band structures of non-stoichiometric surface models. In the case of a Se-rich surface of $Bi_2Se_3$ ($III_{Se}$ termination, Fig. 5a) we observe the presence of topologically trivial states with high band dispersion crossing the band gap and strongly hybridizing with the topologically protected states. The topologically trivial states are distinguished from the topologically protected, the presence of which is guaranteed by the strong topological nature $Bi_2Se_3$ and $Bi_2Te_3$, by means of the bulk-boundary correspondence. The bands corresponding to the trivial states cross some energy within the band gap an even number of times, while an odd number of crossings can be observed for the topologically protected states. In the case of a Bi-rich surface of $Bi_2Se_3$ ($II_{Bi}$ termination, Fig. 5b) the Fermi level appears shifted into the conduction band, i.e. *n*-type doping of the bulk-like states is observed. At the same time, the Dirac point of the topological surface-state band at the $\Gamma$ point appears to be immersed in the valence band. The change in the position of the Dirac point at $\Gamma$ pulls another topologically protected crossing at the time-reversal invariant momentum point $Y$ into the bulk band gap. A similar reorganization of the band dispersion of the topological surface states upon changing the surface structure has been predicted for a model bulk topological insulator.[46] An overview of the general electronic structure features of non-stoichiometric surfaces of $Bi_2Se_3$ and $Bi_2Te_3$ are given in Table 1, while the corresponding band structures are reproduced in Figs. S2 and S3 of the Supplementary Information. In general, we find that *n*-type doping is a characteristic feature of all Bi-rich surfaces, while the presence of trivial mid-gap states is observed for both chalcogen and Bi-rich surfaces in the extreme limits of chemical potential $\mu_{Bi}$.

## Discussion

In summary, our work addresses the atomic structure and electronic properties of high-index surfaces in nanostructures of bismuth chalcogenide topological insulators, $Bi_2Se_3$ and $Bi_2Te_3$, through first-principles calculations. We predict that several possible quintuple-layer terminations of different stoichiometric compositions can be realized, depending on experimental conditions. Both the stoichiometry of the surface and its crystallographic orientation significantly affect the electronic properties of topologically protected surface states, particularly the anisotropy of their Dirac fermion band dispersion and the degree of spin polarization. Moreover, these properties are shown to display clear dependence on the surface configuration. Through this understanding, one can gain a greater degree of control over the properties of nanostructures of topological insulators aiming at prospective technological applications of these novel materials.

## Methods

First principles calculations were performed within the DFT framework, employing the generalized gradient approximation (GGA) to the exchange-correlation functional.[47] Spin-orbit effects were treated self-consistently using fully relativistic norm-conserving pseudopotentials[48] within the two-component wavefunction formalism. A plane-wave kinetic energy cutoff of 35 Ry has been employed for the wave functions. Calculations were implemented through the PWSCF plane-wave pseudopotential code of the Quantum-ESPRESSO distribution.[49] Quintuple-layer termination energies at reference chemical potential $G(\mu_{Bi}=0)$ were calculated using single-QL nanoribbon models of 5.4 nm and 5.7 nm width for $Bi_2Se_3$ and $Bi_2Te_3$, respectively. In total, over 60 different termination structures were considered, shown as grey lines in Figs. 2a,b. The electronic structure of high-index surfaces was investigated using two-dimensional slab models of the same thickness as the width of nanoribbon models. The lattice constants of slab configurations were calculated assuming bulk lattice constants as well as the stacking order of bulk crystals.

## Acknowledgements

We thank G. Autès for discussions and technical assistance. This work was supported by the Swiss National Science Foundation (Grants No. PP00P2_133552), ERC project "TopoMat" (grant No. 306504) and by the Swiss National Supercomputing Centre (CSCS) under project s515.

## Author contributions

O.V.Y. conceived the ideas and supervised the project. N. V. performed calculations and analysed the data. Both authors wrote the paper.

## Additional information

**Supplementary information** accompanies this paper at http://…

**Competing financial interests:** The authors declare no competing financial interests.

**Figures**

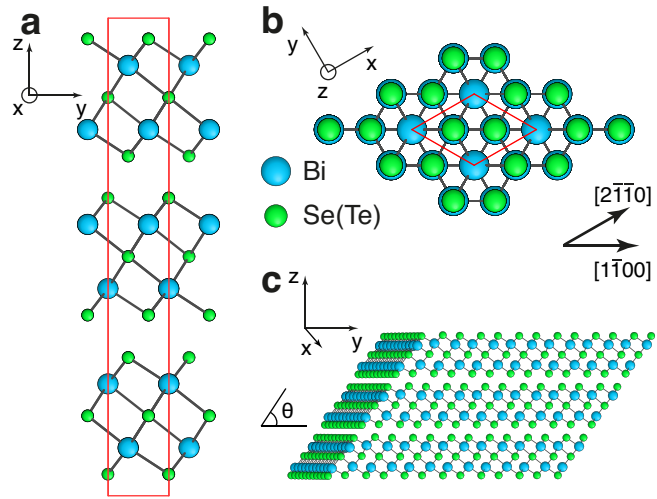

**Figure 1 | Atomic structure of bulk bismuth chalcogenides and their high-index surfaces.** (a), (b) Crystal structure of bulk $Bi_2Se_3$ and $Bi_2Te_3$ viewed along the *a* and *c* axes, respectively. The hexagonal unit cell is shown using red lines. High-symmetry crystallographic directions $[2\bar{1}\bar{1}0]$ and $[1\bar{1}00]$ are indicated in panel (b). (c) Two-dimensional slab model illustrating the structure of one example of high-index surface (left facet). The structure of high-index surfaces of layered bismuth chalcogenides is defined by stacking angle $\theta$ and the QL termination. The example shown corresponds to $\theta = 57.7°$ and the stoichiometric QL termination (configuration I).

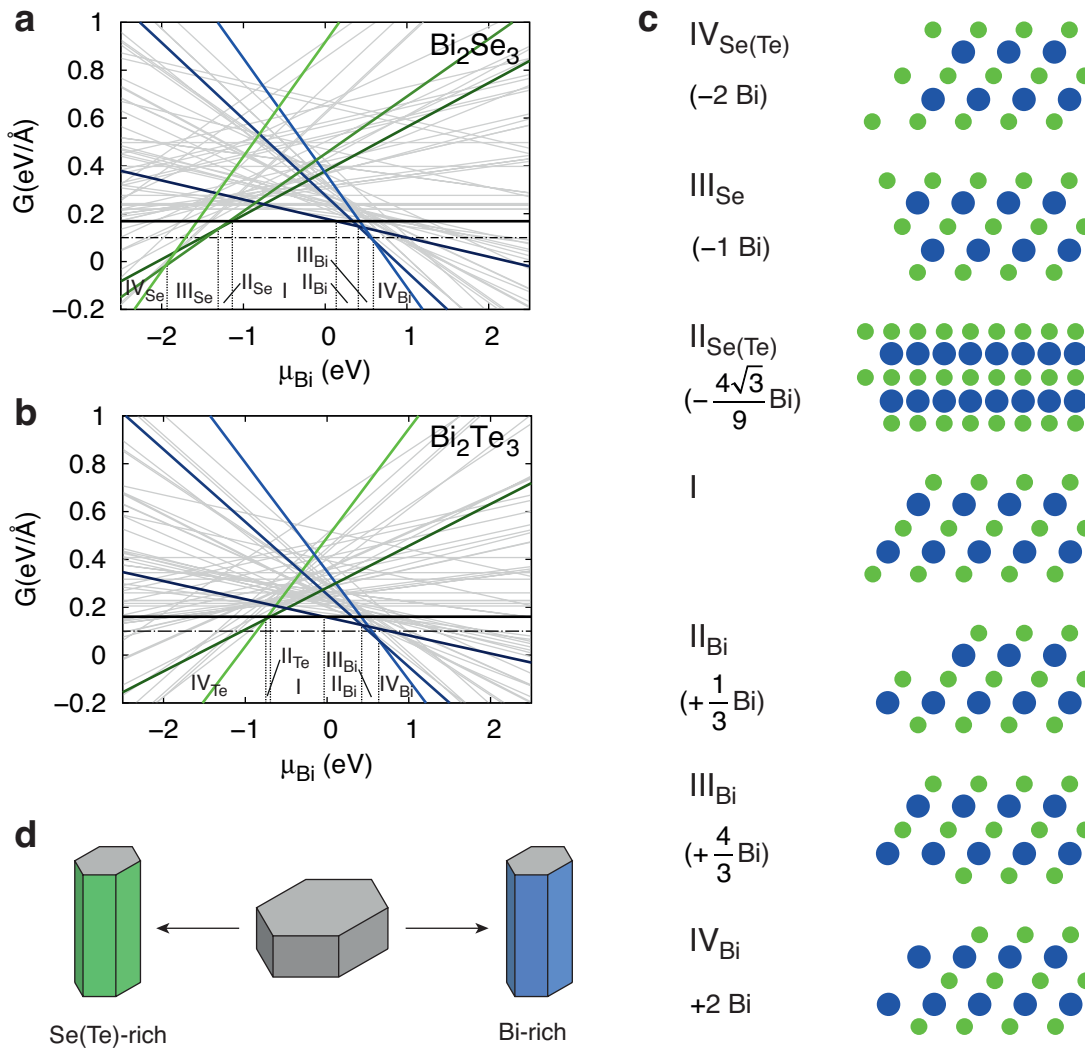

**Figure 2 | Structure and energetics of QL terminations.** (a), (b) QL termination free energies $G(\mu_{Bi})$ of $Bi_2Se_3$ and $Bi_2Te_3$ QLs, respectively, calculated for a large number of structural models. Coloured lines indicate termination structures showing regions of stability in certain ranges of chemical potential $\mu_{Bi}$. (c) Atomic structures of stable QL terminations. Unrelaxed structures are shown for clarity. For non-stoichiometric terminations, the deviations from the nominal ratio Bi : X = 2 : 3 (in Bi atoms per lattice constant *a*) are given in parentheses. (d) Schematic illustration showing that at moderate values of $\mu_{Bi}$ calculations predict the formation of nanoplates, whilst at extreme positive and negative values of $\mu_{Bi}$ formation of nanowires is predicted, where the dominating surfaces are either Bi- or Se(Te)-rich, respectively.

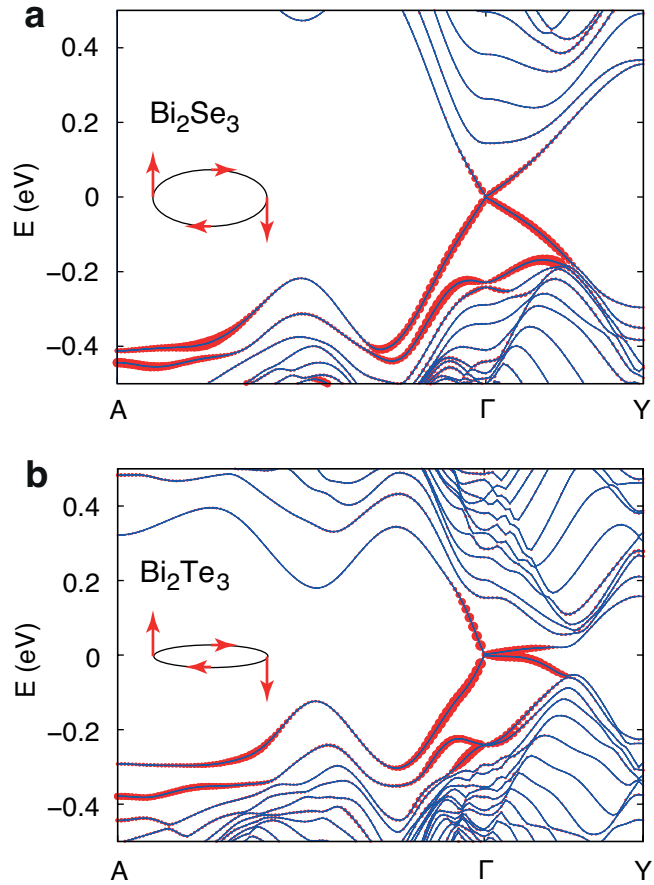

**Figure 3 | Band structures of stoichiometric high-index slab models of $Bi_2Se_3$ and $Bi_2Te_3$.** First-principles band structures of slab models of stoichiometric (configuration I) high-index surfaces of (a) $Bi_2Se_3$ and (b) $Bi_2Te_3$. The atomic structure of the surfaces is shown in Fig. 1c and corresponds to $\theta = 57.7°$ and $\theta = 58.0°$ for $Bi_2Se_3$ and $Bi_2Te_3$, respectively. Points *A* and *Y* correspond to the Brillouin zone boundaries along directions defined by reciprocal lattice vectors associated with the real-space unit vectors of the surface. The size of red symbols reflects the magnitude of the inverse participation ratio (IPR). Schematic drawings of the constant energy contours depicting the helicity of electron charge carriers are shown in the insets.

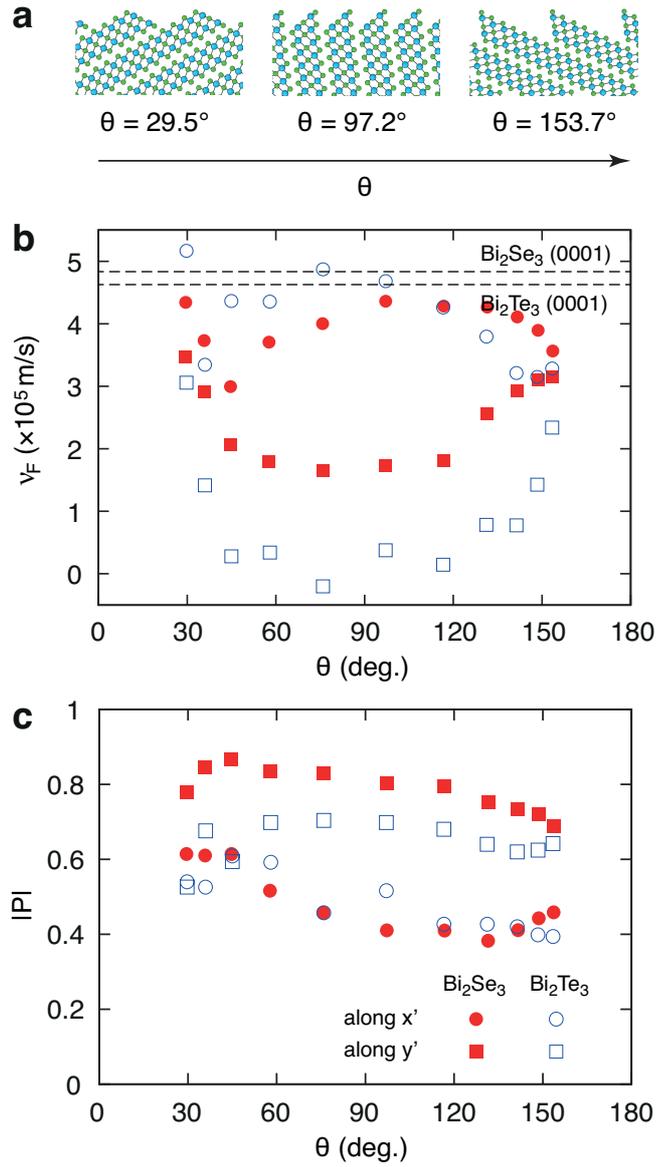

**Figure 4 | Dependence of the band dispersion of topologically protected states on the surface orientation.** (a) Atomic structures of three high-index stoichiometric surfaces of bismuth chalcogenides corresponding to $\theta$ = 29.5°, 97.2° and 153.7°, where $\theta$ defines the QL stacking angle of a high-index surface (see Figure 1c and text). (b) Fermi velocities $v_F$ and (c) absolute values of spin polarization **P** of electron surface-state charge carriers with momenta along *x'* and *y'* directions as a function of surface orientation $\theta$. Values of the Fermi velocity $v_F$ for the (0001) surfaces of $Bi_2Se_3$ and $Bi_2Te_3$ are indicated by dashed lines in panel (b).

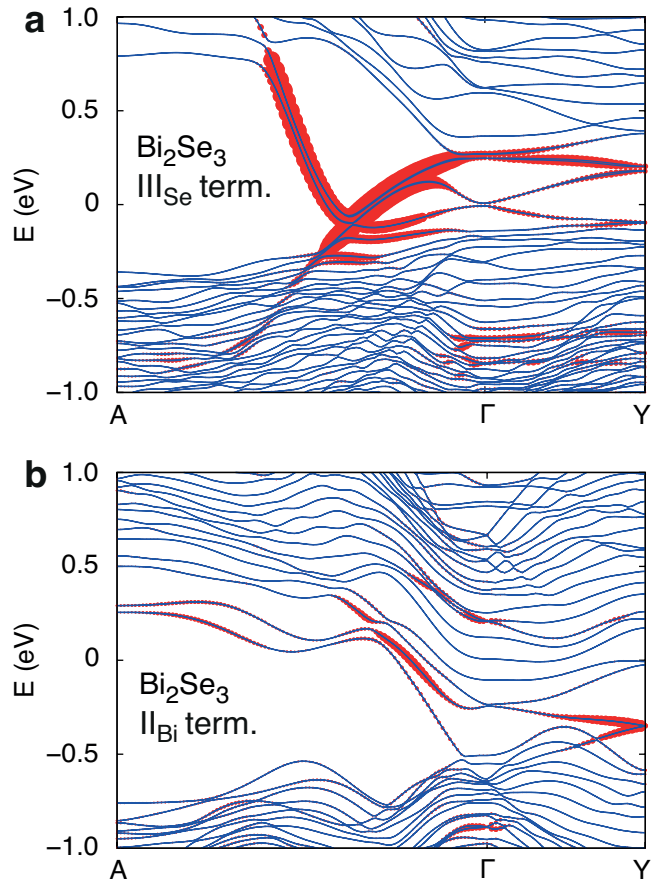

**Figure 5 | Representative examples of band structures of non-stoichiometric high-index slab models.** First-principles band structures of slab models of (a) Se-rich (III$_{Se}$ QL termination) and (b) Bi-rich (II$_{Bi}$ QL termination) high-index surfaces of Bi$_2$Se$_3$ at $\theta = 57.7°$. Points *A* and *Y* correspond to the Brillouin zone boundaries along directions defined by reciprocal lattice vectors associated with the real-space unit vectors of the surface. The size of red symbols reflects the magnitude of the inverse participation ratio (IPR).

# Tables

**Table 1 | Summary of electronic properties of bismuth chalcogenide surfaces with various stoichiometric compositions. The presence of topologically trivial mid-gap states and self-doping are pointed out.**

|  | QL termination | trivial mid-gap states | self-doping |
|---|---|---|---|
| Bi$_2$Se$_3$ | IV$_{Se}$ | Yes | - |
|  | III$_{Se}$ | Yes | - |
|  | II$_{Se}$ | No | - |
|  | I | No | - |
|  | II$_{Bi}$ | No | $n$-type |
|  | III$_{Bi}$ | No | $n$-type |
|  | IV$_{Bi}$ | Yes | $n$-type |
| Bi$_2$Te$_3$ | IV$_{Te}$ | Yes | - |
|  | II$_{Te}$ | No | - |
|  | I | No | - |
|  | II$_{Bi}$ | No | $n$-type |
|  | III$_{Bi}$ | No | $n$-type |
|  | IV$_{Bi}$ | Yes | $n$-type |